\documentclass{aa}
\usepackage{psfig}
\usepackage{latexsym}

\newcommand{\omcen}{$\omega\,$Cen}

\newcommand{\x}{X$\,$}
\newcommand{\ltap}{\mathrel{\hbox{\rlap{\lower.55ex \hbox {$\sim$}}
                   \kern-.3em \raise.4ex \hbox{$<$}}}}
\newcommand{\gtap}{\mathrel{\hbox{\rlap{\lower.55ex \hbox {$\sim$}}
                   \kern-.3em \raise.4ex \hbox{$>$}}}}
\newcommand{\nh}{N_{\rm H}}
\newcommand{\cmsq}{{\rm cm}^{-2}}
\newcommand{\ctks}{cts\,ksec$^{-1}$}
\newcommand{\ergs}{erg\,s$^{-1}$}
\newcommand{\ergcms}{{\rm erg\,cm}^{-2}{\rm s}^{-1}}

\begin{document}
%%%%%%%%%%%%%%%%%%%%%%%%%%%%%%%%%%%%%%%%%%%%%%%%%%%%%%%%%%%%%%%%%%%%% TITLE
\thesaurus{05(10.07.3 $\omega$\,Cen, NGC6397, NGC6752, Liller\,1, 13.25.5)}
\title{Multiple and variable X-ray sources in the globular clusters
\omcen, NGC\,6397, NGC\,6752, and Liller~1}
%\subtitle{      }
\author{F.~ Verbunt\inst{1} \and H.M.~Johnston\inst{2}
 }
\offprints{F.~Verbunt}
\mail{verbunt@phys.uu.nl}

\institute{     Astronomical Institute,
              P.O.Box 80000, NL-3508 TA Utrecht, The Netherlands
         \and              
             Anglo Australian Observatory, P.O. Box 296, Epping,
             NSW 2121, Australia
                        }
\date{\today, accepted}   
\authorrunning{F.\ Verbunt \&\ H.M.\ Johnston}
\titlerunning{Dim X-ray sources in four globular clusters}
\maketitle

%%%%%%%%%%%%%%%%%%%%%%%%%%%%%%%%%%%%%%%%%%%%%%%%%%%%%%%%%%%%%%%%%% ABSTRACT

\begin{abstract}
We detect stars from the Hipparcos and Tycho Catalogues in the field of
view of observations  with the ROSAT HRI of three globular clusters.
We use the positions of these stars to reduce the systematic error
in the positions of X-ray sources in the clusters to $\sim2''$
for \omcen\ and NGC\,6752, and $1''$ for NGC\,6397.
We detect three X-ray sources in the core of \omcen, and four in the
core of NGC\,6752; the data for the center of NGC\,6397 may be fitted with
five or six sources.
Outside the cores, but within the half-mass radius of the
clusters, we detect two sources in \omcen, one in NGC\,6397 and two in
NGC\,6752; these may or may not be cluster members.
A ROSAT HRI observation of Liller\,1 does not detect a low-luminosity
source, at a limit below a detection with ASCA.
We discuss the nature of the low-luminosity X-ray sources in globular
clusters in the light of these new results.
\keywords{globular clusters: individual $\omega$\,Cen, NGC\,6397,
NGC\,6752, Liller\,1 -- X-rays: stars}
\end{abstract}

\section{Introduction}

Globular clusters contain many X-ray sources at lower luminosities,
$L_{\rm x}\ltap 10^{34}$\,\ergs.
These sources were first discovered with the Einstein satellite
(Hertz \&\ Grindlay 1983), and many more were found with ROSAT
(for a compilation, see Johnston \&\ Verbunt 1996).
The nature of these low-luminosity sources is the subject of debate,
because various types of objects can emit X-rays at such luminosities,
such as soft X-ray transients in quiescence, cataclysmic variables,
RS~CVn binaries, and recycled neutron stars (see e.g.\ Fig.~8 in
Verbunt et al.\ 1997).
The most compelling identification of a dim X-ray source with an object 
observed at other wavelengths is the recycled radio pulsar in M\,28:
the X-ray flux varies on the pulse period (Danner et al.\ 1994).
Plausible identifications with cataclysmic variables have been
suggested for dim X-ray sources in NGC\,6397, NGC\,6752, NGC\,5904
and 47\,Tuc (Cool et al.\ 1995b, Grindlay 1993, Hakala et al.\ 1997,
Verbunt \&\ Hasinger 1998).
These identifications are based on the proximity of the X-ray position to
that of a cataclysmic variable, and thus their probability depends on the
accuracy of the X-ray position.
\nocite{hg83}\nocite{jv96}\nocite{vbrp97}\nocite{dkt94}\nocite{cgc+95}
\nocite{gri93}\nocite{hcjv97}\nocite{vh98}

The accuracy of the ROSAT position of a detected X-ray source 
is determined by two factors: the statistical accuracy of the position
of the source on the detector, and the accuracy with which the position
of the detector as a whole is projected on the sky.
For a sufficient number of photons the statistical error is less than
an arcsecond, but the projection in general has a typical error of $\sim5''$.
Secure identification of a source in the detector field reduces the
error in the projection to the statistical
error of the identified source, provided that the optical (or radio)
position has better accuracy. Only one identification is necessary,
because the roll angle of the detector (i.e.\ the North-South direction)
is accurately known; nonetheless, identification of more than one source
is preferable to allow checks on internal consistency.
In a globular cluster the surface density of possible counterparts is so high 
that chance coincidence usually cannot be excluded; a secure identification 
can usually be made only for X-ray sources detected well outside the cluster.
This method has been used by Verbunt \&\ Hasinger (1998) to improve
the positional accuracy of the sources in the core of 47\,Tuc from
$5''$ to $2''$, whereby the area in which the source is expected to
lie is reduced sufficiently to exclude several proposed identifications,
and increase the probability of others, including two possible cataclysmic 
variables.

In this paper we investigate three clusters known to contain
multiple dim X-ray sources in their core, which have been observed
in long exposures with the ROSAT HRI, and one cluster known to harbour
a transient.
We analyse hitherto unpublished observations and detect both previously
published and new X-ray sources. 
All source positions are checked in the SIMBAD data base versus positions
of other objects, and we find objects in the Hipparcos or Tycho Catalogues
(ESA 1997, Perryman et al.\ 1997, H{\o}g et al.\ 1997) 
with each cluster, i.e.\ counterparts with very accurate positions.
In Sect.\,2 we describe the observations and our data reduction procedures;
Sections 3 to 5 describe the results for \object{$\omega$\,Cen}, 
\object{NGC\,6397}, and \object{NGC\,6752}, respectively. In Sect.\,6 
we discuss an observation of \object{Liller\,1}. 
A discussion of our results is given in Sect.\,7.
\nocite{ESA97}\nocite{plk+97}\nocite{hbb+97}

\section{Observations and data reduction}

\begin{table}
\caption[o]{Log of the ROSAT HRI observations of globular clusters
analysed in this paper. For each cluster observation,
the observation date(s) and exposure time are given. 
We further give the shift in $\alpha,\delta$
applied to bring the X-ray coordinate frame of 
the longest observation to the optical coordinate frame J2000.
\label{ta}}
\begin{tabular}{l@{\hspace{0.2cm}}ll@{\hspace{0.2cm}}rl@{\hspace{0.2cm}}l}
& \multicolumn{2}{c}{\mbox observing period} & $t_{\rm exp}$(s) 
& $\Delta\alpha$ & $\Delta\delta$ \\
\multicolumn{5}{c}{\omcen} \\
1992 & Aug & 2448835.935--837.027 & 3469  &  & \\
1993 & Jan & 2449008.688--008.702 & 1243  \\
1994 & Jul & 2449547.617--550.695 & 5997  \\
1995 & Jan & 2449735.780--744.786 & 17653  \\
1995 & Jul & 2449912.812--941.103 & 75481 & $-0\fs05$ & $0\farcs0$ \\
1996 & Feb & 2450120.247--130.418 & 12990  \\
1997 & Jan & 2450461.343--461.358 & 1317 \\
\multicolumn{5}{c}{NGC\,6397} \\
1991 & Mar & 2448338.820--340.054 & 7499 \\
1992 & Mar & 2448696.631--697.646 & 12958 \\
1995 & Mar & 2449793.872--810.440 & 77204 & $+0\fs11$ & $-0\farcs4$ \\
\multicolumn{5}{c}{NGC\,6752} \\
1992 & Mar & 2448697.719--703.626 & 31300 & $-0\fs66$ & $-1\farcs3$ \\
1995 & Mar & 2449800.941--815.431 & 17411 & \\
1996 & Apr & 2450182.957--186.041 & 23278 & \\
\multicolumn{5}{c}{Liller\,1} \\
1996 & Sep & 2450332.541--337.810 & 16412  \\
\end{tabular}
\end{table}

The X-ray observations were obtained with the ROSAT X-ray telescope
(Tr\"umper et al.\ 1991) in combination with the high-resolution imager
(HRI, David et al.\ 1995).
The list of the observations is given in Table~\ref{ta}.
The standard data reduction was done with the Extended Scientific Analysis 
System (Zimmermann et al.\ 1996), as follows.
To take into account the re-calibration of the pixel size (Hasinger et 
al.\ 1998), we multiply the $x,y$ pixel coordinates of each photon with
respect to the HRI center with 0.9972. 
A search for sources is made by comparing counts in a box with
the counts in a ring surrounding it, and by moving this detection
box across the image. 
The sources thus detected are excised from the image and a background map
is made for the remaining photons.
A search for sources is then made by comparing the number of photons in
a moving box with respect to the number expected on the basis of the
background map.
Finally, at each position in which a source was found, 
a maximum-likelihood technique is used to compare the observed
photon distribution with the point spread function of the HRI
(Cruddace et al.\ 1988). 
This produces a maximum-likelihood value ML such that the probability
that the source is due to chance at one trial position is given by
$e^{-{\rm ML}}$.
We retain sources for further discussion if ML$\geq13$.
(To make sure that all such sources are found, we enter in the maximum
likelihood technique all sources
that have ML$\geq$10 according to the sliding box searches.)

Upper limits at the position of known sources were determined by
counting the number $n$ of actually detected photons at the position of
the source (and in an area surrounding it corresponding to the
uncertainty in the position); we then assign as upper limit the lowest
expected number $m$ for which the probability of measuring a number $n$ or 
smaller is less than 5\%\ according to the Poisson distribution.
\nocite{tha+91}\nocite{dhkz95}\nocite{zbb+96}\nocite{hbg+98}\nocite{chs88}

The maximum-likelihood technique also provides an indication whether
the source is extended. If such indication is present, we apply further 
analysis to test whether the source is a multiple point source.

The further analysis is also based on maximum likelihood techniques,
but the analysis is limited to a small area of the detector, near its
center. This allows the simplifications that the background in the
analysed area is a constant (as opposed to a polynomial function
of the $x,y$ pixel coordinates), and that the point spread function is that
for the center of the image (David et al.\ 1995).
Suppose that a model to be tested predicts $m_i$ photons at detector pixel $i$.
The probability that $n_i$ photons are observed is then given by the
Poisson probability:
\begin{equation}
P_{i} = {{m_i}^{n_i}e^{-m_i}\over n_i!}
\end{equation}
The probability that the model describes the observations is
given by the product of the probabilities for all $i$ in the region considered:
$L'=\Pi P_i$. 
For computational ease we maximize the logarithm of this quantity:
\begin{equation}
\ln L' \equiv \sum_{i} \ln P_{i} = \sum_{i}n_i\ln m_i -\sum_{i}m_i 
 -\sum_{i}\ln n_i!
\end{equation}
The last term in this equation doesn't depend on the assumed model,
and -- in terms of selecting the best model -- may be considered as a
constant. Thus maximizing $L'$ is equivalent to minimizing $L$, where 
\begin{equation}
\ln L \equiv -2(\sum_{i}n_i\ln m_i -\sum_{i}m_i )
\end{equation}
If one compares two models A and B, with number of fitted parameters
$n_A$ and $n_B$ and with likelihoods of $\ln L_A$ and $\ln L_B$, respectively,
the difference $\Delta L\equiv\ln L_A-\ln L_B$ is a $\chi^2$ distribution with
$n_A-n_B$ degrees of freedom, for a sufficient number of photons
(Cash 1979, Mattox et al.\ 1996).
\nocite{cas79}\nocite{mbc+96}

Our analysis of possibly multiple sources thus proceeds as follows.
First we compute $\Delta L$ for a model with constant background and
for the best model with background plus one source, and compare it
with the $\chi^2$-distribution with 3 degrees of freedom.
If $\Delta L>15$, the presence of one source has a significance
more than three sigma.
Next we compare the best model with two sources with the best model
with one source, to prove the significance of a second source;
the best models with three and two sources to prove the significance
of a third source, etc.\ until no more significant sources are found.

The addition of one source adds three fitted parameters, one for its 
number of counts and two for its position.
In the case of NGC\,6397 optical counterparts have been suggested
for three X-ray sources. For these we also make a fit in which
the distances in right ascension and declination between these
three sources is fixed to the optically determined values. The three
sources in that case only add five fitted parameters, two for the
position of one of them, and three for the fluxes.

\begin{figure}
%\centerline{\psfig{figure=omcen.ps,width=0.7\columnwidth,clip=t} {\hfil}}
\centerline{\psfig{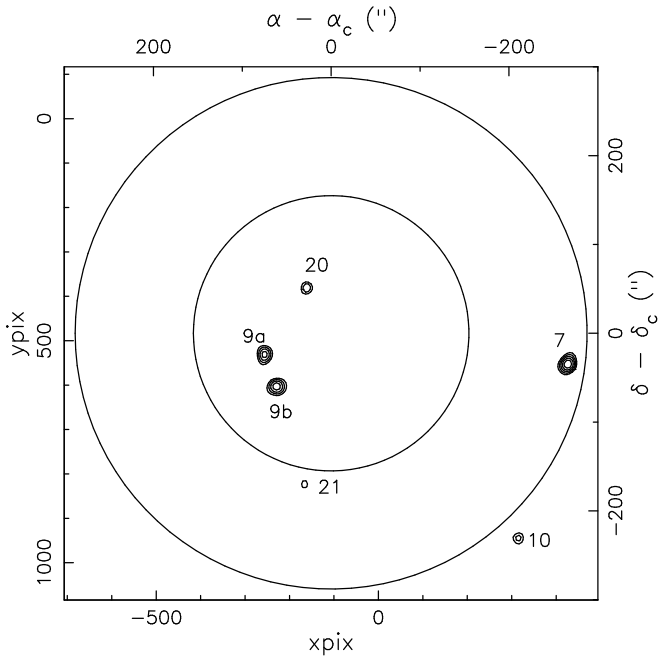} {\hfil}}
\caption{X-ray contours in the central area of \omcen\ as observed with 
the ROSAT HRI in 1995 July.
The image was smoothed with a 2-d $\sigma$$\simeq$5$''$ Gaussian.
The detected sources are indicated with their numbers in Table\,\ref{tabomc}.
The inner circle gives the core radius of the cluster, the outer circle the
half-mass radius.
The lower and left axes give pixel numbers for the ROSAT HRI detector,
the upper and right axes right ascension and declination with respect
to the cluster center. The conversion between pixel and celestial coordinates
is accurate to within 2$''$.
\label{figomc}}
\end{figure}
\nocite{tdk93}

To determine the error in a parameter, we start from the best fit value $a_i$.
We then fix the parameter at $a_i+d$ and make a new fit, allowing all other
parameters to vary. The value of $d$ for which $\ln L$ increases by 1 is quoted
as the 1-sigma error.

\section{\omcen}

\omcen\ is a very massive globular cluster, with a relatively
low central density. 
Hertz \&\ Grindlay (1983) reported five sources \object{A-E near \omcen}.
Sources A, D and E all are well outside the cluster core, and it appears
that only source\,C is clearly related to the globular cluster
(e.g.\ Verbunt et al.\ 1995).
Source\,E has tentatively been identified with a foreground K star
(Margon \&\ Bolte 1987).
The sources\,A and D have been identified with foreground
M stars, on the basis of better positions for the X-ray sources
obtained with Einstein and ROSAT HRI observations (Cool et al.\ 1995a). 
A ROSAT PSPC pointing indicates that source\,C, near the cluster center,
is composed of two sources of comparable luminosity (Johnston et al.\ 1994).
\nocite{vbhj95}\nocite{mb87}\nocite{jvh94}

\begin{table}
\caption{X-ray sources detected in the globular cluster \omcen\ ($A_V=0.47$,
$d=4.9$\,kpc, Djorgovski 1993) with the
ROSAT HRI. Sources with a number less than 17 correspond to sources
detected previously with the ROSAT PSPC (Johnston et al.\ 1994); those with
a higher number are new sources. For each source we give the position,
the statistical error in the position (in $''$), 
the distance to the cluster center
in units of the core radius $r_c$, the countrate with error, and where
applicable the identification with sources detected with Einstein 
(Hertz \&\ Grindlay 1983). The sources are ordered on declination.
The positions given are those after correction for bore sight (see text).
The positions of the center of the cluster (GC, Djorgovski \&\ Meylan 1993), 
its core radius and half-mass radius (Trager et al.\ 1993), and positions of
some optical objects discussed in the text are also
listed; epochs are 1990.5 for positions by Cool et al., 1979 for USNO-A2,
and 1995.5 for HD\,116789.
\label{tabomc}}
\begin{tabular}{rr@{ }r@{ }rr@{ }r@{ }rccr@{ }l}
\multicolumn{7}{c}{X-ray sources} \\
X  &  \multicolumn{3}{c}{$\alpha$\,(2000)}
   &  \multicolumn{3}{c}{$\delta$\,(2000)} & $\Delta$ & $d/r_c$ &
cts/ksec  \\ 
4 &  13 &  27 &  27.72 & $-$47 &  19 &   8.0 & 0.4 &   4.6 &   7.8$\pm$0.4 &D\\
3 &  13 &  25 &  52.05 & $-$47 &  19 &   8.6 & 0.5 &   5.1 &   6.3$\pm$0.3&A \\
18 &  13 &  26 &  46.33 & $-$47 &  19 &  45.9 & 1.0 &   3.4 &   0.5$\pm$0.1 \\
19 &  13 &  27 &  21.21 & $-$47 &  23 &  22.0 & 1.8 &   3.1 &   0.4$\pm$0.1 \\
6 &  13 &  27 &  29.27 & $-$47 &  25 &  54.5 & 1.3 &   3.0 &   0.9$\pm$0.2 \\
20 &  13 &  26 &  48.70 & $-$47 &  27 &  46.5 & 1.0 &   0.4 &   0.6$\pm$0.1 \\
9a &  13 &  26 &  53.39 & $-$47 &  29 &   1.5 & 0.9 &   0.5 &   0.9$\pm$0.1&C\\
7 &  13 &  26 &  19.75 & $-$47 &  29 &  11.9 & 0.8 &   1.7 &   1.4$\pm$0.2&B \\
9b &  13 &  26 &  52.04 & $-$47 &  29 &  37.2 & 0.9 &   0.6 &   1.1$\pm$0.2&C \\
21 &  13 &  26 &  48.99 & $-$47 &  31 &  28.1 & 2.1 &   1.1 &   0.5$\pm$0.1 \\
10 &  13 &  26 &  25.20 & $-$47 &  32 &  28.9 & 1.4 &   2.0 &   0.7$\pm$0.1 \\
13 &  13 &  26 &  11.12 & $-$47 &  37 &  11.1 & 3.7 &   4.0 &   1.3$\pm$0.3 \\
\multicolumn{7}{c}{optical objects} \\
GC & 13 & 26 & 45.9 & $-$47 & 28 & 37 & 
\multicolumn{4}{l}{$r_c=155''$, $r_h=288''$} \\
4  &  13 &  27 &  27.37 & $-$47 &  19 &   6.2 & 
   \multicolumn{4}{l}{Cool et al.\ (1995a)} \\
4  &  13 &  27 &  27.68 & $-$47 &  19 &   6.3 &
    \multicolumn{4}{l}{USNO-A2\,0375-18249604} \\
3  &  13 &  25 &  51.79 & $-$47 &  19 &   7.0 &
   \multicolumn{4}{l}{Cool et al.\ (1995a)} \\
3  &  13 &  25 &  52.23 & $-$47 &  19 &   7.5 &
    \multicolumn{4}{l}{USNO-A2\,0375-18177834} \\
18 &  13 &  26 &  46.33 & $-$47 &  19 &  45.9  &
\multicolumn{4}{l}{HD\,116789} \\
E  &  13 &  29 &  18.54 & $-$47 &  22 &  50.5 &
    \multicolumn{4}{l}{Margon \&\ Bolte (1987)} \\
\end{tabular}
\end{table}
\nocite{djo93}\nocite{cgb+95}

We have analysed all observations listed in Table~\ref{ta}. 
The 1992 and 1993 data have been reported on before by Cool et al.\ (1995a).
As expected, we detect the largest number of sources in the longest 
observation, that of July 1995. We use this observation as the basis 
for comparison with the other observations.

\subsection{Source list and membership probability}

\begin{figure*}
%\centerline{\psfig{figure=omcenlt.ps,width=15.cm,clip=t}}
\centerline{\psfig{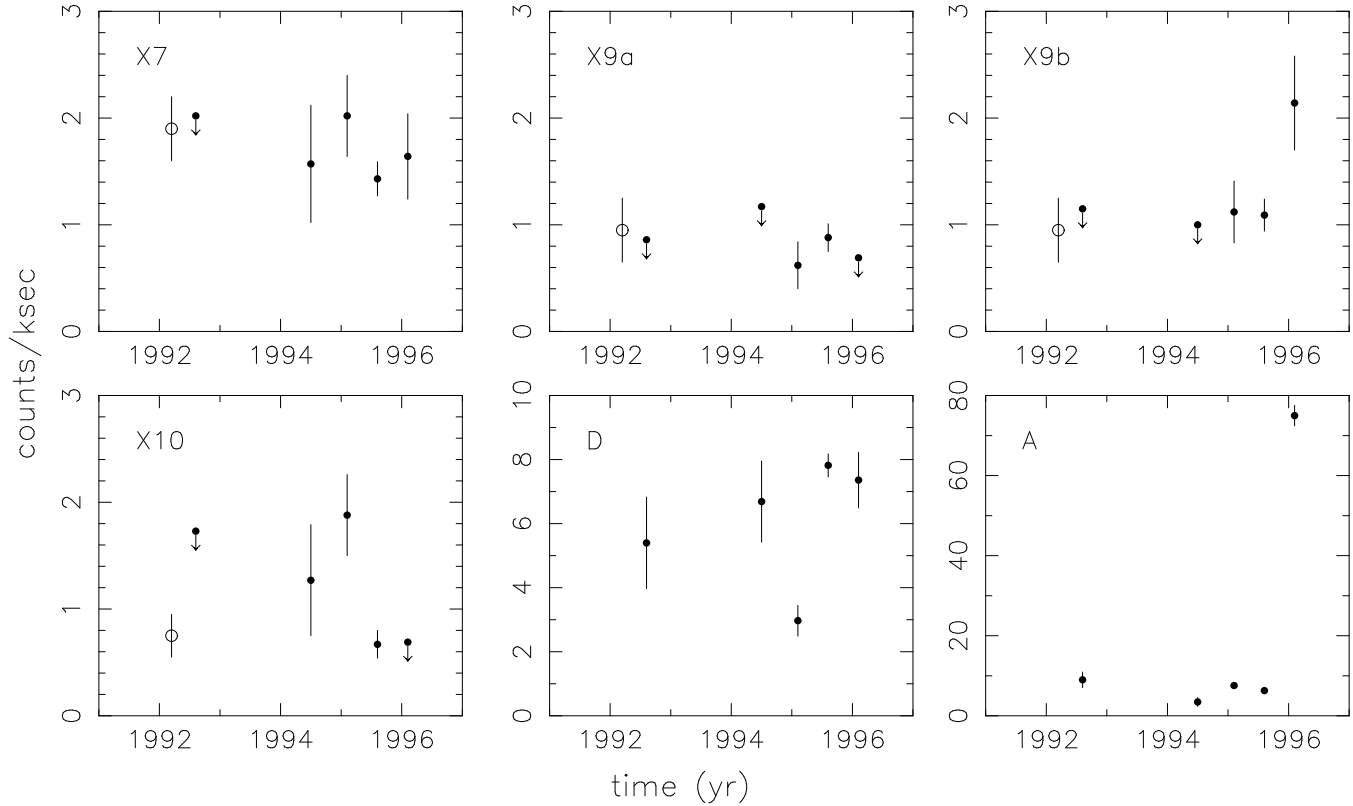} {\hfil}}
\caption{X-ray lightcurves of four sources in the center of \omcen\ and
of two foreground M dwarfs (A, D). The HRI data are shown as $\bullet$ 
(detections, with 1-$\sigma$ errors), $\downarrow$ (2-$\sigma$ upper limits). 
PSPC data (converted to estimated HRI countrates) are shown as $\circ$.
There is marginal evidence for variability in sources \x9b and \x10; the
foreground sources A and D are highly variable, with source A showing a
large flare in 1996.
\label{figlt}}
\end{figure*}

In the July 1995 observation we detect twelve sources, listed in 
Table\,\ref{tabomc}. Eight of these sources have been detected before
with Einstein or with the ROSAT PSPC observation reported by
Johnston et al.\ (1994); four sources are new.

The sources \x3 and \x4 are the Einstein sources A and D, respectively,
identified with foreground M dwarfs by Cool et al.\ (1995a, see 
Table\,\ref{tabomc}).
Both stars can be found in the USNO-A2 catalogue (Monet et al.\ 1998,
see Table\,\ref{tabomc}).
The new source \x18 can be identified with \object{HD\,116789}; this is star 
TYC\,8252\,4627\,1 in the Tycho Reference Catalogue, and thus its position 
and proper motion are well known (H{\o}g et al.\ 1998). 
For this reason we use this star to determine the bore sight correction, 
i.e.\ the offset between the X-ray coordinates and the optical coordinates. 
The result is listed in Table\,\ref{ta}.
\nocite{mbc+98}\nocite{hkb+98}

This bore sight correction has been applied to the X-ray positions, and the 
resulting positions are given in Table\,\ref{tabomc}.
With a statistical accuracy for the X-ray position of \x18 of $\sim1$$''$,
and taking into account small additional systematic errors (see Hasinger et 
al.\ 1998), we estimate the systematic error in the
positions given in Table\,\ref{tabomc} to be less than $2''$; for each
individual source its statistical uncertainty should be added in quadrature
to this systematic error.
The positional accuracy can be improved if accurate astrometry of
\x3/A and \x4/D is obtained, which will allow computation of their
positions at epoch 1995.5.

In the ROSAT Deep Survey (Hasinger et al.\ 1998), 
an area with 12$\farcm$5 radius contains
25 sources brighter than our approximate detection limit of
$0.7\times10^{-14}\,\ergcms$; we thus expect $\sim$1 serendipitous source
within the core radius, i.e.\ the faintest source within the core
radius, \x20, may well be a serendipitous background source.
The brighter sources within the core radius probably are associated with 
\omcen.
Outside the core radius a source is more likely to be a
fore- or background source than a cluster member.
\nocite{hbg+98}

\subsection{Sources in the cluster}

The X-ray image of the inner area of \omcen\ is shown in Fig.\,\ref{figomc}.
The half-mass radius of the cluster contains five sources.
Source \x9 from Johnston et al.\ (1994) is clearly separated into two
sources, which we denote \x9a and \x9b for the northern and southern source,
respectively. 

We have determined the countrates or upper limits for the six central 
sources in 
all ROSAT HRI observations of \omcen\ listed in Table\,\ref{ta}. For the
very short observations, no useful upper limits are obtained; the
long-term lightcurves as determined from the other observations are shown
for four of the central sources in Fig.\,\ref{figlt}.
We also show the PSPC observation, dividing the PSPC counts for 
PSPC-\x9 equally between \x9a and \x9b. For the absorption towards \omcen\ and
a black body spectrum of 0.6\,keV
the PSPC countrate is about 2.8 times the HRI countrate; for smaller
reddening the PSPC-to-HRI count ratio varies rapidly, and therefore we do not
show PSPC points of the foreground M dwarfs, whose absorption is unknown.

There is marginal evidence for variability
in sources \x9b and \x10; and no evidence for variability of \x7 and \x9a.
\x20 is detected only in July 1995 and in 1996, \x21 only in July 1995,
and all upper limits in the other observations are compatible with the
faint fluxes of these sources.

The July 1995 observation was obtained in two parts, separated by
about 10 days. We have analysed the two parts separately, and
find marginal variation of \x7 ($1.7\pm0.3$ and $1.1\pm0.2$ \ctks\ in the 
first and second half, respectively) and of \x9b
($0.8\pm0.2$ and $1.4\pm0.2$ \ctks, respectively). In addition,
virtually all counts of \x21 are from the first part of the observation.
The number of counts of \x21 is too small for further subdivision.

To convert the observed countrates into X-ray luminosities, we 
assume a column and distance to \omcen\ as given in Table\,\ref{tabomc}.
For an 0.6\,keV blackbody (see the analysis of the PSPC spectrum of
\x9 in Johnston et al.\ 1994) 1 \ctks\ in the ROSAT HRI corresponds to
$1.5\times10^{32}$\,\ergs\ in the 0.5--2.5\,keV band.
The two sources \x9a and \x9b thus have about this luminosity,
source \x7 is 40\%\ brighter, and sources \x20 and \x21 are
40\%\ fainter.

\subsection{Sources not related to the cluster}

\begin{figure}
%\centerline{\psfig{figure=flare.ps,width=0.7\columnwidth,clip=t} {\hfil}}
\centerline{\psfig{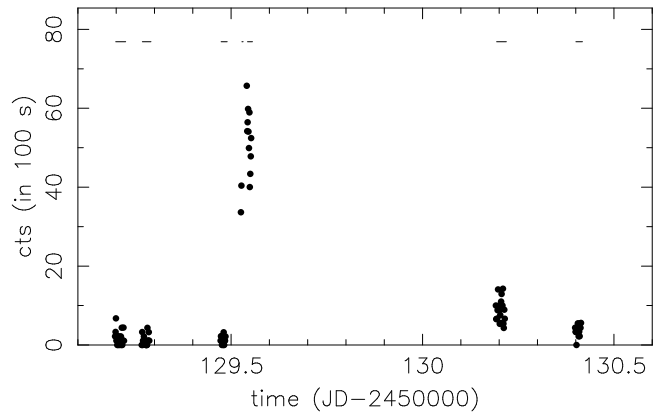} {\hfil}}
\caption{Lightcurve of \x3/A in the 1996 observation. $\bullet$ indicate
the numbers of counts collected in 100\,s intervals; the horizontal
lines indicate when ROSAT was collecting data. 
\label{flare}}
\end{figure}

The foreground dwarfs (Einstein X-ray sources A and D) are highly
variable, as has been pointed out before (Koch-Miramond \&\
Auri\`ere 1987, Cool et al.\, 1995a). These sources also vary between
the first and second half of the July 1995 observation. 
The extremely high flux of A in 1996 is due to a flare, which lasts
almost a day (see Fig.\,\ref{flare}).
\nocite{ka87}

HD\,116789 is an A0V star. From its magnitude $V=8.40$ and colour
$B-V=0.07$ we estimate $A_V=0.22$ and a distance of about 310\,pc. 
For an assumed bremsstrahlung spectrum of 1.4\,keV 
the observed countrate corresponds to a luminosity
in the 0.5--2.5\,keV band of $\sim2\times10^{29}$\,\ergs.
It is not expected for an A0V star to emit such a flux; perhaps this
star has a white dwarf companion which is responsible for the
X-ray emission, as various other A0V stars detected with ROSAT;
on the other hand, various apparently single
A0V stars in the Bright Star Catalogue 
have been detected at similar and higher luminosities as HD\,116789
(e.g.\ HD\,17864, H\"unsch et al.\ 1998).
HD\,116789 was detected with EXOSAT by Verbunt et al.\ (1986), who 
interpreted the detection as due to the ultraviolet leak of the CMA detector.
The EXOSAT CMA countrate of this source, $1.7\times10^{-3}$\,cts\,s$^{-1}$,
converts to an X-ray luminosity at the distance of HD\,116789 of
about $\sim10^{31}$\,\ergs, much higher than the luminosity derived
for this star from the ROSAT observations; we therefore still think
that the EXOSAT countrate is due to the ultraviolet flux.
The ultraviolet leak in the ROSAT HRI is far too small (Bergh\"ofer et 
al.\ 1999) to explain the ROSAT detection.
\nocite{hsv98}\nocite{vsj+86}\nocite{bsh99}

\subsection{Discussion}

The positions of the sources within the half-mass radius of
\omcen\ as given in Table\,\ref{tabomc} are more accurate than
previously published positions, and may be used to search for
optical counterparts.
We have done this among the variables (contact binaries, detached binaries, 
and suspected RS CVn stars) found in \omcen\ by Kaluzny et 
al.\ (1996, 1997): no counterpart is among these stars.
(Only one of these variables is in the area shown in Fig.\,\ref{figomc}, 
viz.\ the contact binary OGLEGC\,13.)
Our non-detection of these binaries is not surprising, considering that
our detection limit is above $10^{31}$\,\ergs: all of the
contact binaries hitherto detected in X-rays (McGale et al.\ 1996), 
and many RS CVn systems (Dempsey et al.\ 1993) are less
luminous than this.
\nocite{kks+96}\nocite{kks+97}\nocite{mph96}\nocite{dlfs93}

Cool et al.\ (1995a) argue that \x7/B is an extended source.
This source is detected in the ROSAT PSPC observation (Johnston et 
al.\ 1994) and in the ROSAT HRI observations of 1994, 1995 January and
July, and 1996. All of these observations are more sensitive than
the 1992 and 1993 observations used by Cool et al. (1995a); in all of
them \x7/B is compatible with being a point source.

With the identification of sources \x3 and \x4 with foreground
stars, we can reinvestigate the suggested identification of 
\x5/E with a foreground K star, as suggested by Margon \&\ Bolte
(1987). 
We use the optical positions of A and D to determine the
offset between the X-ray positions of the PSPC observation
as listed in Johnston et al. (1994) and the optical positions.
We then apply this offset to the position of \x5, and find that
the resulting position is at 6$\pm$5 arcseconds from the
optical star. We take the position of the optical star 
(Table\,\ref{tabomc}) from USNO\,A2\,0375-18334783 (Monet et al.\ 1998).
Identification of
\x5 with the foreground star is therefore a distinct possibility. 

\section{NGC\,6397}

NGC\,6397 is a nearby cluster, with a collapsed core, in or close to which 
Cool et al.\ (1993) detected \object{four X-ray sources (B, C1-3)} with a 
ROSAT HRI observation. Photometry with the Hubble Space Telescope enabled
Cool et al.\ (1995b) to find eight candidate counterparts for these sources,
on the basis of high ultraviolet flux or of H\,$\alpha$ emission.
The H\,$\alpha$ emission of three stars has been confirmed spectroscopically
by Grindlay et al.\ (1995) who argue that these stars are cataclysmic
variables, and responsible for the X-ray emission close to the core.
\nocite{cgkb93}\nocite{cgc+95}\nocite{gcc+95}

\subsection{Source list and membership}

We analyse first the longest observation, obtained in 1995, and use this
as a reference for our discussion of the earlier, shorter observations.
The standard analysis provides 14 sources, listed in Table\,\ref{taba}.
Identifications with earlier X-ray sources or optical objects are
indicated; 7 sources are new.
\x6 has been identified by Cool et al.\ (1993) as SAO\,244944.
This star is identical to \object{HD\,160177}, and is in the Hipparcos
Catalogue as HIP\,86569. Its position and proper motion are thus
very accurately known, and we use it to determine the bore sight correction.
This bore sight correction is given in Table\,\ref{ta}, and is applied to
the X-ray positions; the resulting positions are given in Table\,\ref{taba}.
The statistical uncertainty in the X-ray position of \x6 is about 0.5$''$;
we therefore estimate that systematic error of the X-ray positions listed in 
Table\,\ref{taba} is better than 1$''$; this error should be added in
quadrature to the statistical error for each individual source position.
The quasar identified by Cool et al.\ (1993) with \x5 coincides within the
error with our position for \x5.
However, the active galaxy identified by Cool et al.\ (1993) with \x2 is 
10$''$ from our X-ray position, mainly in right ascension; and we 
conclude that it is not the X-ray source. The explanation probably lies
in the new scale for the size of the HRI pixel that we use (see Sect.\,2),
which modifies positions of sources at large distance from the center of 
the HRI image.

\begin{table}
\caption{X-ray sources detected in the NGC\,6397 ($A_V=0.56$, $d=2.2$\,kpc,
Djorgovski 1993) 
with the ROSAT HRI, for the standard analysis of the whole field, and
separately for two multi-source analyses of the central area.
Numbers up to 10 are sources from Johnston et al.\ (1994), higher numbers are
new; cross-identifications with sources discussed by Cool et al.\ (1993)
are listed on the right.
All X-ray positions have been corrected for boresight.
The positions of the center of the cluster (GC, Djorgovski \&\ Meylan 1993),
its core radius and half-mass radius (Trager et al.\ 1993) and the
positions of some optical objects discussed in the text are also listed; 
epochs are 1992.7 for positions by Cool et al., and 1996.3 for HIP\,86569.
\label{taba}}
\begin{tabular}{rr@{ }r@{ }rr@{ }r@{ }rrrr@{ }l}
X  &  \multicolumn{3}{c}{$\alpha$\,(2000)}
   &  \multicolumn{3}{c}{$\delta$\,(2000)} & $\Delta$ & $d/r_c$ &
cts/ksec \\ 
\multicolumn{11}{c}{X-ray sources in HRI field} \\
2 &  17 &  41 &  34.79 & $-$53 &  32 &   2.0 & 0.8 & 230 &   4.9$\pm$0.3 &G \\
11 &  17 &  40 &  51.16 & $-$53 &  33 &  49.0 & 2.0 & 135 &   0.4$\pm$0.1 \\
12 &  17 &  40 &  48.85 & $-$53 &  39 &  46.0 & 1.2 &  25 &   0.4$\pm$0.1 \\
13$^a$ &  17 &  40 &  41.56 & $-$53 &  40 &   6.0 & 0.5 &   6.3 &   4.6$\pm$0.3 &B \\
4 &  17 &  40 &  42.24 & $-$53 &  40 &  23.8 & 0.4 &   2.2 &   7.8$\pm$0.3 &C\\
14 &  17 &  40 &   1.43 & $-$53 &  42 &  25.6 & 1.6 & 125 &   0.3$\pm$0.1 \\
15 &  17 &  39 &  16.54 & $-$53 &  43 &  11.3 & 3.7 & 258 &   2.8$\pm$0.3 \\
5 &  17 &  40 &  33.18 & $-$53 &  43 &  46.6 & 0.4 &  72 &   3.6$\pm$0.2 &D \\
16 &  17 &  41 &  23.73 & $-$53 &  46 &  17.2 & 1.0 & 172 &   1.2$\pm$0.2 &E \\
17 &  17 &  39 &  32.92 & $-$53 &  46 &  48.7 & 3.5 & 240 &   0.7$\pm$0.2 \\
18 &  17 &  40 &  30.81 & $-$53 &  47 &  51.4 & 0.8 & 152 &   1.1$\pm$0.1 \\
6 &  17 &  41 &  27.66 & $-$53 &  48 &  10.6 & 0.5 & 207 &   5.5$\pm$0.3 &F\\
19 &  17 &  40 &  14.53 & $-$53 &  50 &  31.7 & 2.8 & 217 &   0.6$\pm$0.1 \\
8 &  17 &  40 &  10.46 & $-$53 &  50 &  54.7 & 1.7 & 229 &   1.4$\pm$0.2 \\
\multicolumn{11}{l}{$^a$position affected by nearby source 4} \\
\multicolumn{11}{c}{X-ray sources near center; 5-source fit} \\
13 & 17 & 40 & 41.44 & $-$53 & 40 & 3.3 & 0.3 & & 3.1$\pm$0.2 & B\\
4a & 17 & 40 & 42.47 & $-$53 & 40 & 28.6 & 0.5 & & 2.4$\pm$0.3 \\
4b & 17 & 40 & 42.55 & $-$53 & 40 & 19.1 & 0.5 & & 2.5$\pm$0.2 & ID\,3 \\
4c & 17 & 40 & 41.68 & $-$53 & 40 & 19.0 & 0.9 & & 1.2$\pm$0.2 & ID\,1 \\
4d & 17 & 40 & 41.64 & $-$53 & 40 & 27.7 & 0.7 & & 1.5$\pm$0.3 \\
\multicolumn{11}{c}{X-ray sources near center; 6-source fit$^b$} \\
13 & 17 & 40 & 41.44 & $-$53 & 40 & 3.3 & 0.3 & & 3.1$\pm$0.2 & B \\
4a & 17 & 40 & 42.24 & $-$53 & 40 & 28.6 & 0.4 & & 1.7$\pm$0.3 & ID\,2 \\
4b & 17 & 40 & 42.57 & $-$53 & 40 & 19.3 & & & 2.6$\pm$0.2 & ID\,3 \\
4c & 17 & 40 & 41.52 & $-$53 & 40 & 19.4 & & & 1.1$\pm$0.2 & ID\,1 \\
4d & 17 & 40 & 41.56 & $-$53 & 40 & 27.7 & 0.8 & & 1.2$\pm$0.2 & \\
4e & 17 & 40 & 42.65 & $-$53 & 40 & 27.6 & & & 1.1$\pm$0.3 & ID\,6 \\
\multicolumn{11}{l}{$^b$positions of ID\,1, 3 and 6 fixed relative to ID\,2}\\
\multicolumn{7}{c}{optical objects} \\
GC &  17 &  40 & 41.3  & $-$53 & 40 & 25 & 
\multicolumn{4}{l}{$r_c=3''$, $r_h=174''$} \\
5  &  17 &  40 &  33.4 & $-$53 &  43 &   45. & 
   \multicolumn{4}{l}{D Cool et al.\ (1993)} \\
2$^c$  &  17 &  41 &  35.9 & $-$53 &  32 &  4. &
   \multicolumn{4}{l}{G Cool et al.\ (1993)} \\
6 &  17 &  41 &  28.0  & $-$53 &  48 &  13.  &
   \multicolumn{4}{l}{F Cool et al.\ (1993)} \\
6 &  17 &  41 &  27.66 & $-$53 &  48 &  10.6  &
\multicolumn{4}{l}{HIP\,86569/HD\,160177} \\
\multicolumn{11}{l}{$^c$suggested identification probably wrong}\\
\end{tabular}
\end{table}

The flux detection limit is about $0.8\times10^{-14}\,\ergcms$ outside
the blended central region, similar to that obtained for \omcen.
Analogous to our argument for \omcen, we find that all objects detected
within $0\farcm5$ are probably cluster members, whereas we expect 1.4
background sources within $3'$ from the center of of NGC\,6397;
the sources at $0\farcm5<r<3'$ therefore may be background
sources. We thus cannot decide whether \x12 is a cluster member.
Outside the half-mass radius, the sources are more likely to be background
or foreground sources. \x5, just outside the half-mass radius, is a quasar 
(Cool et al.\ 1993).

\subsection{The central sources}

\begin{figure}
%\centerline{\psfig{figure=ngca.ps,width=0.7\columnwidth,clip=t} {\hfil}}
\centerline{\psfig{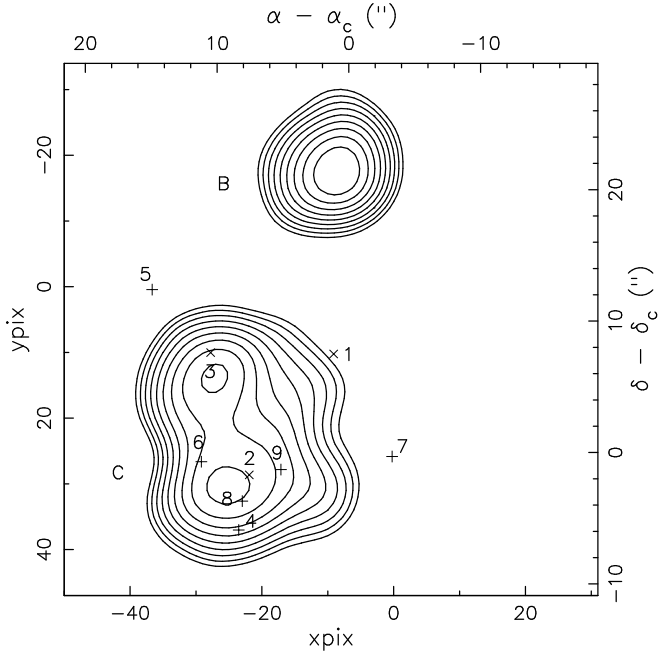} {\hfil}}
\caption{Positions of ultraviolet or H\,$\alpha$-emission objects 
in the central area of NGC\,6397 
($\times$, $+$, numbered with their ID in Table~2 of Cool et al.\ 1995b)
superposed on X-ray contours of sources \x13/B and \x4/C 
as observed with the ROSAT HRI in 1995.
The candidate counterparts for three X-ray sources suggested by Cool et
al.\ are marked $\times$.
The X-ray image was smoothed with a 2-d $\sigma$$\simeq$2$''$ Gaussian.
The lower and left axes give pixel numbers for the ROSAT HRI detector,
the upper and right axes right ascension and declination with respect
to the cluster center. The conversion between pixel and celestial coordinates
is accurate to within 1$''$.
\label{figngca}}
\end{figure}

In Fig.\,\ref{figngca} we show the X-ray contours of the center of NGC\,6397
together with the ultraviolet and/or H\,$\alpha$-emission stars discovered
by Cool et al.\ (1995b). 
The first models we investigated as fits to the central area of 
$60''\times60''$, containing sources \x13/B and \x4/C, are those with 
successively one, two, three, four and five sources; all with free positions. 
Using the $\Delta L$ criterion for significance (see Sect.\ 2) we find that
five sources are required.
We refer to the fit with five sources as Model\,I. The parameters of the 
five sources of this model are given in Table\,\ref{taba}.
We do not detect source A of Cool et al.\ (1993) in the 1995 observation.

\begin{figure}
%\centerline{\psfig{figure=ngcb.ps,width=0.7\columnwidth,clip=t} {\hfil}}
\centerline{\psfig{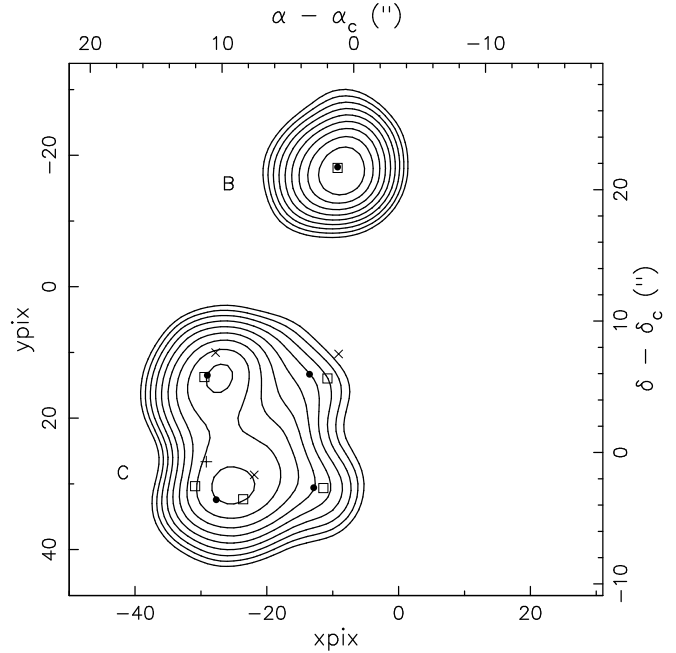} {\hfil}}
\caption{X-ray contours in the central area of NGC\,6397 as observed with 
the ROSAT HRI in 1995, with the positions of the sources obtained
in the best fits. $\bullet$ positions found with the best fit
for four components in C in which all positions are left free (Model\,I),
$\Box$ positions found with the best fit for five components
in C, in which three sources have distances fixed with respect to
one other source, at the distances of ID\,1, 3, and 6 to ID\,2 (Model\,IV).
Other symbols as in Fig.\,\ref{figngca}.
Model\,I is marginally better at the 2-sigma level.
\label{figngcb}}
\end{figure}

Cool et al.\ (1995b) resolved source \x4/C into three components C1-3. 
From a list of H\,$\alpha$ emission and/or ultraviolet excess
objects (their Table~2), they suggest identifications of ID\,1 with C2, 
ID\,2 for C3 and ID\,3 for C1. 
Comparing the positions of the sources in Model\,I we find that the positions
of \x4b and \x4c are compatible with those of ID\,3 and ID\,1, respectively; 
we thus identify \x4b with C1 and \x4c with C2. \x4d is a new source. 
(The offset required to match these positions from Table~\ref{taba} with those
given by Cool et al.\ (1995b) is slightly larger than our claimed accuracy of
$\ltap 1''$; the remaining difference may be explained by an offset between the
Guide Star Catalogue coordinate system and the more accurate Hipparcos
coordinate system.)
The position of \x4a is not compatible with that of ID\,2.
The reason for this may be seen in Fig.\,\ref{figngca}: the two brightest 
components of source \x4/C have a smaller difference in
right ascension than ID\,2 and ID\,3. 
If ID\,2 is the correct identification for C3, we conclude that 
\x4a is not identical to C3.

To further investigate this, we note that if the identifications are correct, 
the distances between the X-ray sources must match the distances between
the proposed optical counterparts, which are accurately known from
the HST observations.
In Model\,II we fit five sources to the X-ray data of the center of NGC\,6397,
of which three are forced to be at fixed relative positions, corresponding
to the distances between ID\,1, ID\,2 and ID\,3.
Model\,II thus has four fitted parameters less than Model\,I.
The $\ln L$ of Model\,II is 26 higher than that of Model\,I,
i.e.\ it is a significantly (4-sigma) worse fit. This confirms that \x4a
is not ID\,2.
In Model\,III we assume that ID\,6 of Table~2
in Cool et al.\ (1995b) rather than ID\,2 is the counterpart of C3,
and fix the distances between the sources accordingly. This
fit has the same $\ln L$ as Model\,II, and thus also is
significantly worse than the fit of Model\,I.
Again, the reason for the bad fit is the mismatch in the difference
in right ascensions of the two brightest X-ray sources with that
between the proposed counterparts: \x4a is not ID\,6.

We note that the best position of \x4a is between ID\,2 and ID\,6, and
in Model\,IV we fit six sources, of which four are forced to be at the
relative distances of  ID\,1-3 and ID\,6.
Model\,IV thus has three fitted parameters less than Model\,I.
Its $\ln L$ is 6 higher than that of Model\,I, i.e.\ it
is marginally worse at less than 2-sigma.
The parameters of the six sources of this model are also
given in Table\,\ref{taba}. It is seen that the positions of \x4b, \x4c 
and \x4d are the same (within the error) in Model\,IV as in Model\,I.

Thus, we have two acceptable models. In both models we confirm the
possible identifications of ID\,3 with C1 ($=$\x4b) and of ID\,1 with
C2 ($=$\x4c), and we find one new source (\x4d).
In Model\,I the remaining flux is ascribed to one source (\x4a) which
is not identical to ID\,2. In Model\,IV the remaining flux is ascribed to
two sources, one of which is ID\,2/C3 and one is a second new source,
\x4e/ID\,6.
The two acceptable models are illustrated in Fig.\,\ref{figngcb}.

\begin{table}
\caption{Results of fitting four models to the three data sets of NGC\,6397.
The Table lists the number $n$ of fitted parameters, and the difference
$\Delta L$ with respect to the best model for a given data set.
In all fits, the fluxes of all sources are fitted parameters.
For the 1995 observation Model\,I has four sources with free positions,
Model\,IV has six sources of which three have free positions and 
three have fixed positions relative to ID\,3, corresponding
to the offsets of ID\,1, ID\,2, and ID\,6 with respect to ID\,3. Models\,II 
and III are as Model\,IV, after removing ID\,6 and ID\,2, respectively.
For each Model, the same positions as in the best fit for 1995 are used for the
1991 and 1992 data.
\label{tabfit}}
\begin{tabular}{l@{\hspace{0.4cm}}rr@{\hspace{0.4cm}}rr@{\hspace{0.4cm}}rr}
   &  \multicolumn{2}{c}{1995} & \multicolumn{2}{c}{1991}
   &  \multicolumn{2}{c}{1992} \\
Model & n & $\Delta L$  & n & $\Delta L$  & n & $\Delta L$ \\
I (B+4)           & 16 & $\equiv$0 & 8 &       1.3 & 8 & 4.7 \\
II (B+1+ID\,123)  & 12 &        26 & 8 &       1.4 & 8 & 7.8 \\
III (B+1+ID\,136) & 12 &        26 & 8 &       3.2 & 8 & $\equiv$0 \\
IV (B+1+ID\,1236) & 13 &       6.2 & 9 & $\equiv$0 & 9 & 1.9 \\
\end{tabular}
\end{table}

\subsection{The earlier observations}

The standard analysis detects \x2, \x5, \x16, \x6 and \x8 in both the 1991
and the 1992 data of NGC\,6397, and \x19 in the 1992 data, all at countrates 
compatible with those of 1995. It also detects sources \x13/B and \x4/C
in the 1991 data and in the 1992 data, labelling both as extended. 
The number of photons in sources B and C is rather small in these short
observations. 
To limit the number of parameters in the fits to the central sources
we demand that the distance between the fitted central sources in each model
is the same as in the best fit to the 1995 data, but allow the fluxes
to be different.
The corresponding reductions in the number of fitted parameters for each model
are indicated in Table\,\ref{tabfit}.

We thus fit four models to each data set. For each year, the best model
is set at $\Delta L\equiv 0$, and the quality of the other models for that
year is determined with respect to this model. 
The results of our fitting are shown in Table\,\ref{tabfit}.
For the 1991 data, the models with five sources are comparable in quality,
and the six-source model is not significantly better.
For the 1992 data, Model\,III is marginally better (2 sigma) than Model\,I
and significantly (3 sigma) better than Model\,II, whereas Model\,IV is 
of similar quality.

\begin{table}
\caption{Countrates (counts\,ksec$^{-1}$)
assigned to the central sources in Models\,I and IV in 
the fits to the observations of 1991 and 1992. Numbers in parentheses
indicate the errors in the last digit. For 1995 see Table\,\ref{taba}.
\label{tabflux}}
\begin{tabular}{lrr@{\hspace{2.cm}}lrr}
  \multicolumn{3}{c}{Model\,I} & 
  \multicolumn{3}{c}{Model\,IV} \\
X & 1991 & 1992 & X & 1991 & 1992  \\
13 & 2.3(7) & 2.9(5) & 13 & 2.5(7) & 2.8(6) \\
4a & 1.4(6) & 0.8(4) & 4a & 1.5(8) & 0.7(5) \\ 
4b & 2.2(7) & 2.1(5) & 4b & 2.2(7) & 2.1(6) \\
4c & 1.8(7) & 2.5(5) & 4c & 1.9(7) & 2.3(6) \\
4d & 0.9(6) & 0.9(4) & 4d & 0.7(6) & 0.8(4) \\
   &        &        & 4e & $<0.6$ & 0.4(3) \\
\end{tabular}
\end{table}

The fits to the earlier data confirm the conclusions that we draw on the
basis of the observation of the long observation of 1995.
Model\,I in which source \x4/C is separated into four components at free
positions is acceptable for all three observations.
Model\,II in which \x4/C is separated into four components at fixed relative 
distances of ID\,1-3, is not acceptable for the 1992 data.
Model\,IV in which \x4/C is separated into five components, four of which
correspond to ID\,1, ID\,2, ID\,3 and ID\,6, also is acceptable for all 
observations.
ID\,2 is not required in 1992, and ID\,6 is not required in 1991.
The latter fact explains why ID\,6 is not present in the analysis by
Cool et al.\ (1993) of the 1991 data.
These conclusions are confirmed by the countrates that Model\,IV ascribes
to the different sources, listed in Table\,\ref{tabflux}

To see whether we can confirm the existence of source A of Cool et al.\ (1993)
we have also added the 1991 and 1992 observation (after shifting the 1992
observation by $3\farcs5,1\farcs0$ in $\alpha,\delta$; compatible with
the shift as determined by Cool et al.).
We fit Model~I to the added image, and compare it with the fit in which
a source is added to model~A. We find $\Delta L=10$, which implies that 
source~A is marginally significant at $\sim2.5\sigma$.
The position ($+2\fs4,-14''$ with respect to source B) and countrate 
(0.4 counts/ks) that we find for
source A are compatible with those given by Cool et al.\ (1993).

\subsection{Sources not related to the cluster}

HIP\,86569 is a K1\,IV/V star with $V=9.44$, $B-V=0.90$, and a parallax
of $0.0167(18)''$. Hipparcos discovered that
this star is a close binary (separation $0.19''$) of stars with
Hipparcos magnitudes $H=10.17(9)$ and $10.47(11)$, respectively.
At a distance of 60\,pc the observed ROSAT HRI countrate converts to an X-ray
luminosity in the 0.5-2.5\,keV band of 
$L_{\rm x}\simeq4\times10^{28}$\,\ergs\ (for assumed 1.4\,keV bremsstrahlung
with no absorption). 
This is similar to the X-ray luminosities of single KV stars
detected in the ROSAT All Sky Survey, such as HD\,17925 (K1V) which has
$L_{\rm x}\simeq 1.2\times10^{29}$\,\ergs\ (H\"unsch et al.\ 1998).

\subsection{Discussion}

The core of NGC\,6397 contains at least four X-ray sources detected with
ROSAT, and possibly five. 1 \ctks\ for a source at the distance and with
the absorption column of NGC\,6397, for an assumed 0.6\,keV
bremsstrahlung spectrum corresponds to a luminosity in the
0.5-2.5\,keV band of $2.2\times10^{31}\,$\ergs. The faintest source we
detect, \x4c, is at this level. The brightest source is \x13/B, at 
a luminosity of about $6.8\times10^{31}$\,\ergs. 
These luminosities are at the bright end of the luminosity distribution
for cataclysmic variables, such as the large sample investigated with ROSAT 
(Verbunt et al.\ 1997), as expected for an X-ray selected sample.

Of these sources, \x13 and \x4b have the same flux level in all three
observations. Source \x4c is fainter in 1995.
The identifications of \x4b and \x4c with cataclysmic variables
ID\,3 and ID\,1 remains probable, as does the argument by Edmonds et 
al.\ (1999) that these systems are DQ\,Her type systems.
The distance between \x4b and \x4c in Model\,I is marginally less
than the distance between the proposed counterparts; it is tempting to
speculate that this is due to a small X-ray flux contribution of a fourth
cataclysmic variable (`CV\,4') identified by Cool et al.\ (1998) and confirmed
by Edmonds et al. (1999). 
\nocite{cgc+98}\nocite{egc+99}

The new source \x4d has been detected in the 1995 observation because of
the longer exposure; it may, but need not, be brighter in the 1995 observation
than in 1991 and 1992.

If the remaining flux is assigned to one source \x4a, then this source is not
identified, and was brighter in 1995 than in 1992.
If the remaining flux is distributed over two sources \x4a and \x4e, the
flux of \x4a may still be constant, and \x4a may be identified with the
probably DQ~Her type cataclysmic variable ID\,2.
In this case, the flux of \x4e has increased between 1991 and 1995.
ID\,6 was reported to vary by 1.1 magnitude in five hours by De~Marchi 
\&\ Paresce (1994), but was constant in a ten hour observation by Cool et 
al.\ (1998). It is suggested by Edmonds et al.\ (1999) that ID\,6 is a
undermassive helium white dwarf, probably in a binary.
If it is a single helium white dwarf, it cannot be a variable X-ray source;
if it is in a binary with a recycled radio pulsar, it also is unlikely
to be a variable X-ray source; if it is in a binary with another white dwarf,
then optical and X-ray variability can be due to variable mass transfer
from that other white dwarf. 
However, it is also possible that not ID\,6, but a nearby hitherto
unidentified star in NGC\,6397 is the X-ray source \x4e.
\nocite{mp94}

Whether source \x4a alone, or source \x4a and \x4e are present in the
core of NGC\,6397, and in the latter case whether \x4e is identical to
ID\,6 requires a better spatial resolution for the X-ray observations
than  provided by ROSAT.

\section{NGC\,6752}

Two sources have been detected in the core and two near the core
of NGC\,6752 in a ROSAT HRI observation obtained in 1992 (Grindlay 1993); 
close to one of the core sources,
two candidate cataclysmic variables have been identified on the basis
of H\,$\alpha$ emission and variability on (presumably orbital) periods
of 5.1 and 3.7 hrs (Bailyn et al.\ 1996).
\nocite{brs+96}

\begin{table}
\caption{X-ray sources detected in NGC\,6752 ($A_V=0.12$, $d=4.2$\,kpc,
Djorgovski 1993) with the ROSAT HRI, for the standard analysis of the whole 
field, and separately for the multi-source analysis of the central area.
Numbers up to 15 are sources from Johnston et al.\ (1994), higher numbers are
new.
All X-ray positions have been corrected for boresight.
Identifications with letter on the right refer to Grindlay (1993).
The positions of the center of the cluster (GC, Djorgovski \&\ Meylan 1993),
it core radius and half-mass radius (Trager et al.\ 1993) and the positions
of some optical objects discussed in the text are also listed.
\label{tabb}}
\begin{tabular}{rr@{ }r@{ }rr@{ }r@{ }rrrr@{ }l}
X  &  \multicolumn{3}{c}{$\alpha$\,(2000)}
   &  \multicolumn{3}{c}{$\delta$\,(2000)} & $\Delta$ & $d/r_c$ &
cts/ksec  \\ 
\multicolumn{11}{c}{X-ray sources in HRI field} \\
3 &  19 &  12 &  27.01 & $-$59 &  48 &  20.1 & 2.9 &  87 &   3.5$\pm$0.4 \\
13 &  19 &   9 &  59.88 & $-$59 &  54 &  50.6 & 1.1 &  42 &   1.0$\pm$0.2 \\
4  &  19 &  10 &   3.26 & $-$59 &  55 &  33.9 & 0.5 &  38 &   3.1$\pm$0.2 \\
6  &  19 &  10 &  40.23 & $-$59 &  58 &  39.3 & 0.9 &   8.4 &   1.0$\pm$0.1&A \\
7  &  19 &  10 &  51.35 & $-$59 &  59 &   3.0 & 0.6 &   1.0 &   3.6$\pm$0.3&BC \\
14 &  19 &  10 &  55.73 & $-$59 &  59 &  35.9 & 1.1 &   4.4 &   0.7$\pm$0.1&D \\
15 &  19 &  10 &   4.25 & $-$60 &   2 &  54.3 & 0.7 &  73 &   1.8$\pm$0.2 \\
16 &  19 &  10 &   4.56 & $-$60 &   3 &  16.2 & 0.8 &  70 &   1.5$\pm$0.2 \\
17 &  19 &  11 &  20.33 & $-$60 &   3 &  18.5 & 1.2 &  70 &   0.5$\pm$0.1 \\
18 &  19 &  10 &  32.92 & $-$60 &   3 &  59.8 & 1.3 &  67 &   0.5$\pm$0.1 \\
19 &  19 &  11 &  41.66 & $-$60 &   5 &   9.2 & 2.2 &  75 &   0.6$\pm$0.1 \\
20 &  19 &  10 &  12.07 & $-$60 &   6 &   6.6 & 1.8 &  67 &   0.6$\pm$0.1 \\
11$^a$ &  19 &  10 &  57.91 & $-$60 &  16 &  16.5 & 0.9 &  62 &  26.1$\pm$0.7 \\
\multicolumn{11}{l}{$^a$position affected by nearby detector edge} \\
\multicolumn{11}{c}{X-ray sources near center} \\
7a & 19 & 10 & 51.43 & $-59$ & 58 & 56.6 & 0.5 & & 1.6$\pm$0.2 & C \\
7b  & 19 & 10 & 51.20 & $-59$ & 59 &  8.2 & 0.6 & & 1.7$\pm$0.2 & B \\
21 & 19 & 10 & 52.51 & $-59$ & 59 &  1.9 & 1.1 & & 0.5$\pm$0.1 \\
22 & 19 & 10 & 51.34 & $-59$ & 59 & 24.0 & 1.5 & & 0.4$\pm$0.1  \\
\multicolumn{7}{c}{optical objects} \\
GC &  19 &  10 & 51.8  & $-$59 & 58 & 55. & 
\multicolumn{4}{l}{$r_c=11''$, $r_h=115''$} \\
19 &  19 &  11 &  41.66 & $-$60 & 05 & 9.2 & 
   \multicolumn{4}{l}{TYC 9071\,228\,1} \\
11 &  19 & 10 & 57.84 & $-$60 & 16 & 19.1 &
   \multicolumn{4}{l}{HIP\,94235/HD\,178085} \\
16 & 19 & 10 & 04.51 & -60 & 03 & 18.4 &
\multicolumn{4}{l}{USNO-A2\,0225-29896802} \\
   & 19 & 10 & 51.27 & $-$59 & 58 & 53 &
   \multicolumn{4}{l}{\#\,1 Bailyn e.a.\ 1996}\\
   & 19 & 10 & 51.18 & $-$59 & 58 & 49 &
   \multicolumn{4}{l}{\#\,2 Bailyn e.a.\ 1996}\\
\end{tabular}
\end{table}

\subsection{Data analysis and source list}

Two more observations of NGC\,6752 have been obtained by us. Because the
three observations have comparable length, we add them into a combined
image which we analyse and use as a reference for the individual
observations.
To add the three observations we use the method outlined by Verbunt \&\ 
Hasinger (1998), as follows. 
First we correct the data for each observation separately for the
changed pixel size (see Sect.\,2), analyse the resulting images
and determine the
offsets between sources detected in separate observations. Averaging
these offsets we find (on the basis of sources \x3, \x13, \x4, and \x6) 
that the X-ray coordinates of the 1996 observation have to be
shifted by $d\alpha=-0\farcs7\pm0\farcs7$ in right ascension and
$d\delta=+3\farcs6\pm0\farcs7$ in declination to be brought in line with the 
1992 data. Similarly, the 1995 data (on the basis of the same sources plus 
\x14) must be shifted by $d\alpha=+2\farcs4\pm0\farcs7$, 
$d\delta=+1\farcs2\pm0\farcs7$. We apply these corrections to
the pixel coordinates of the photons, and then add the three images
into a combined image, which is analysed in the standard way.
The resulting list of sources in given in Table\,\ref{tabb}.

\begin{figure}
%\centerline{\psfig{figure=ngccall.ps,width=0.7\columnwidth,clip=t} {\hfil}}
\centerline{\psfig{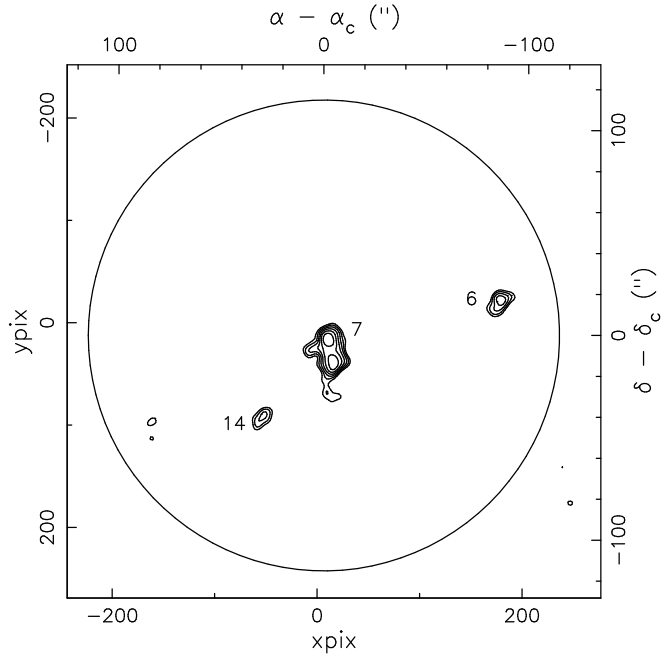} {\hfil}}
\caption{X-ray contours in the central area of NGC\,6752 as observed with 
the ROSAT HRI in the combined image of the 1992-1996 data.
The image was smoothed with a 2-d $\sigma$$\simeq$3$''$ Gaussian.
The detected sources are indicated with their numbers in Table\,\ref{tabomc}.
The circle gives the half-mass radius of the cluster.
The lower and left axes give pixel numbers for the ROSAT HRI detector,
the upper and right axes right ascension and declination with respect
to the cluster center.
\label{figngccall}}
\end{figure}

We identify two sources with stars with accurate positions: 
\x19 is close to  \object{TYC 9071\,228\,1} (CD-60\,7128), a star with $V=9.99$
and $B=10.48$, and \x11 is close to HIP\,94235/\object{HD\,178085}, a G0V star
with $V=8.38$ and $B=9.00$. The latter identification was suggested
already by Johnston et al.\ (1994) on the basis of the PSPC observation.
The chance probability of finding a counterpart at these optical brightnesses
is small, and we consider both identifications secure.
The X-ray position of \x11 is affected by its proximity to the edge of
the HRI detector. For this reason we use \x19 to tie the X-ray to the
optical coordinates.
\x19 is not found in any of the three individual observations, showing
up only in the combined frame. It is a relatively weak source and its
position accordingly has an error of $1\farcs5$ both in right ascension
and in declination. The shift required to bring \x19 in coincidence with
TYC 9071\,228\,1 is given in Table\,\ref{ta}, and has been applied to
all positions of the X-ray sources; the resulting positions are listed
in Table\,\ref{tabb}.
(The remaining offset between \x11 and HIP\,94235 is within the nominal error
for the right ascension, and within 2-sigma for the declination: note that
the error is composed of the statistical uncertainties in the positions of 
both \x19 and \x11.)

The detection limit in the total observation is about 
$0.7\times10^{-14}\,\ergcms$. 
An area with radius $12\farcm5$ in the ROSAT Deep Survey contains 25 sources
brighter than this limit; we thus expect to find 0.6 in the
region within the half-mass radius of NGC\,6752, $r_h\simeq2'$.
The sources in the core thus probably belong to the cluster, and possibly the
two sources \x6/A and \x14/D as well.

\subsection{Sources in the center of the cluster}

\begin{figure}
%\centerline{\psfig{figure=ngcc.ps,width=0.7\columnwidth,clip=t} {\hfil}}
\centerline{\psfig{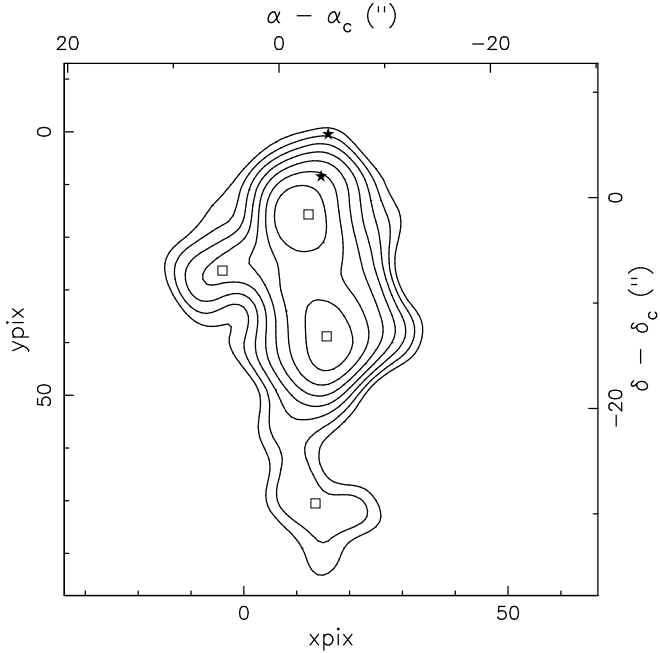} {\hfil}}
\caption{X-ray contours in the central area of NGC\,6752 as observed with 
the ROSAT HRI in the combined data of 1992-1996, with the positions of the 
sources obtained with the multi-source fit ($\Box$).
The positions of two candidate cataclysmic variables
found by Bailyn et al.\ (1996) are indicated with $\star$.
\label{figngcc}}
\end{figure}

Analysing the central source with the method described in Sect.\,2,
we find four significant sources (the increase in $\ln L$ is 29 both
for the third and for the fourth source). 
This adds two sources to the two already described by Grindlay (1993).
The position and fluxes of these sources are listed in Table~\ref{tabb};
Fig.\,\ref{figngcc} shows the positions and X-ray contours of the center
of NGC\,6752, together with the positions of the two candidate
cataclysmic variables found by Bailyn et al.\ (1996). 
The southern cataclysmic variable (`star 1') is at $3.8\pm2.3''$ from \x7a,
and therefore remains a possible counterpart (assuming an error of
1$''$ for the optical position, and taking into account the 2$''$ error
of \x19).

We have analysed the separate observations, keeping the position of the
four central sources fixed at those of the co-added image (as listed in
Table\,\ref{tabb}), but allowing their fluxes to vary. We do not
find significant evidence for variation; due to the limited statistics
we cannot exclude variations by a factor two.

A countrate of 1 \ctks\ corresponds to a luminosity between 0.5 and 2.5
keV of $4.6\times 10^{31}$\,\ergs\ at the distance of NGC\,6752 and for an 
assumed 0.6\,keV bremsstrahlung spectrum.
Thus, \x7a and \x7b have X-ray luminosities of about $7.5\times10^{31}$\,\ergs,
and \x21 and \x22 about a quarter of this.
\x6 and \x14, the two sources outside the cluster core have luminosities 
of $4.6\times 10^{31}$\,\ergs\ and $3.2\times 10^{31}$\,\ergs, respectively,
if they are in NGC\,6752.

\subsection{Sources not related to the cluster}

The spectral type of TYC\,9071\,228\,1 is not known; on the basis of
its magnitude and colour ($V=9.99$, $B-V=0.5$) the star could be a late
F star at a distance of $\sim154$\,pc. At this distance and for
an assumed unabsorbed 1.4\,keV bremsstrahlung spectrum, the
countrate of \x19 converts to an X-ray luminosity in the 0.5-2.5\,keV band 
of $\sim3\times10^{28}$\,\ergs,
a reasonable value for a late F main-sequence star (see e.g.\ the list
of ROSAT detections of bright stars by H\"unsch et al.\ 1998).
The Tycho Catalogue marks this star with 'unresolved duplicity',
with visual magnitude varying between 9.51 and 10.88.

HIP\,94235 has a significant parallax which puts it at
57\,pc. Its countrate converts to an X-ray luminosity at that distance of
about $2\times10^{29}$\,\ergs, a normal X-ray luminosity for a G0V star.

Comparison of the ROSAT image with the USNO-A2 Catalogue gives a candidate
identification for \x16, at a distance of $2\farcs2$, see Table\,\ref{tabb}.
No other sources outside the cluster have been identified by us.

We have analysed the three separate HRI observations, and find no
evidence for variablility, except for \x15, which in March 1992 had an
X-ray flux about half of that observed in March 1995 and April 1996. 

\section{Liller 1}

The globular cluster Liller\,1 is a highly reddened cluster near the
galactic center ($A_{\rm V}\simeq9.5$, $d=8.6\,$kpc, Frogel et al.\ 1995).
It  probably has undergone core collapse (Djorgovski 1993).
Liller\,1 harbours the \object{Rapid Burster}, a highly unusual recurrent 
transient.
When discovered in 1977 the source emitted short ($\ltap5$\,s) bursts of X-rays
every $\sim10$\,s; in some later observations, e.g.\ Aug 1985, it emitted
bursts of $\sim500$\,s separated by 1500-4000\,s; and it has also been
observed as a steady source. The bursts are interpreted as accretion
events. In addition to these, thermonuclear bursts have also been
detected, identifying the accreting star as a neutron star.
A review of this remarkable source is given by Lewin et al.\ (1995).
A low-luminosity X-ray source near Liller\,1 is tentatively identified as the 
quiescent (low-state) counterpart of the Rapid Burster (Asai et al.\ 1996).
\nocite{djo93}\nocite{fkt95}\nocite{lpt95}\nocite{adkk96}

No source is detected in the cluster in our ROSAT HRI observation of
the globular cluster Liller~1.
Near the cluster center, no circle with radius of 5$\arcsec$
contains more than 4 photons. For an expected number of 10 photons,
the probability of getting 4 or fewer photons is less than 4\%.
We thus take 10 as the 2-$\sigma$ upper limit to the number of
photons, which with the effective exposure time is converted to
an upper limit of 0.6 \ctks. 

Asai et al.\ (1996) report the detection on 1993 Aug 27 with ASCA of a 
source near Liller\,1. For a powerlaw with photon index 2,
absorbed by a column $\nh=10^{22}\cmsq$, this source has an unabsorbed flux 
in the 2--10 keV band of $2.5^{+1.7}_{-0.8}\times10^{-13}\,\ergcms$.
For this spectrum our upper limit in the ROSAT HRI
corresponds to a flux of $1.4\times10^{-13}\,\ergcms$, slightly
lower than the ASCA detection.

The ROSAT HRI detects a source with a countrate of 1 \ctks\ about 4$'$
from the cluster center.
The statistical error in the position of this source is about 1$''$; the actual
error is dominated by the error in the bore sight correction,
which is about 5$''$. The ROSAT source is not compatible with the
center of Liller\,1, and also not compatible with the position
of the Rapid Burster as determined with Einstein (see Table\,\ref{tabpos}).
The position of the ROSAT source coincides within the bore sight uncertainty
with the O4 III(f) star \object{HD\,317889} (Vijapurkar \&\ Drilling 1993).
The star is in the Tycho Catalogue as TYC\,7380\,976\,1.
From the observed magnitude and colours ($V=10.12$, $B-V=0.92$,
$U-B=-0.23$, Drilling 1991) we estimate a reddening and distance of 
$E(B-V)\simeq1.2$ and $d\simeq3$\,kpc for HD\,317889. The observed
ROSAT HRI countrate is as expected for such a star, according to
the general correlation between bolometric luminosity and X-ray
luminosity of O stars: $L_x\simeq10^{-7}L_b$ (e.g.\ Kudritzki et al.\ 1996).
(HD\,317888 is within 1$''$ of the O4 star; we have not been able to 
find more information on this star.)
\nocite{vd93}\nocite{dri91}\nocite{kpf+96}\nocite{pj95}\nocite{hg83}

\begin{table}
\caption{Positions of the center of Liller\,1 (GC, Picard \&\ Johnston 1995)
its core radius (Trager et al.\ 1993) and the positions of X-ray sources 
detected near it, viz.\ the Rapid Burster (RB, Hertz \&\ Grindlay 1983), a dim
source detected with ASCA (A, Asai et al.\ 1996) and a dim source
detected with the ROSAT HRI (R, this paper). The position of the O star
HD\,317889 is also given.
The final columns gives the errors in the positions.
\label{tabpos}}
\begin{tabular}{lr@{ }r@{ }rr@{ }r@{ }rrr}
   &  \multicolumn{3}{c}{$\alpha$\,(2000)}
   &  \multicolumn{3}{c}{$\delta$\,(2000)} & $\delta\alpha('')$ &
$\delta\delta('')$ \\
GC & 17 & 33 & 24.5 & $-$33 & 23 & 20.4 & 
\multicolumn{2}{l}{$r_c=4''$} \\
RB & 17 & 33 & 24.0 & $-$33 & 23 & 16.2 & 2 & 2 \\
A  & 17 & 33 & 19.0 & $-$33 & 23 &      &  90 & 30\\
A$^a$ & 17 & 33 & 24.2 & $-$33 & 23 & 6.5 & 20 & 20 \\
R  & 17 & 33 & 4.76 & $-$33 & 23 & 27.2 &  5 & 5\\
HD & 17 & 33 & 5.02 & $-$33 & 23 & 28.4 \\
\end{tabular}

$^a$new determination by Asai, 1999, private communication
\end{table}

\begin{figure}
%\centerline{\psfig{figure=fxlum.ps,width=0.7\columnwidth,clip=t} {\hfil}}
\centerline{\psfig{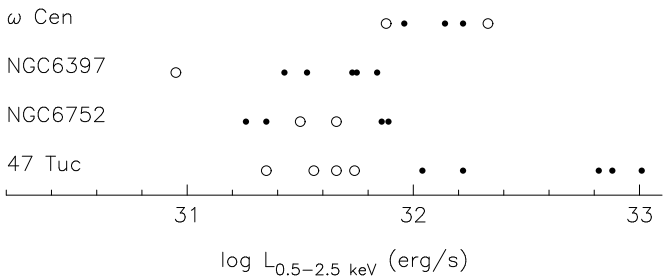} {\hfil}}
\caption{X-ray luminosities of dim sources in four globular clusters.
Sources in and outside the core are shown as $\bullet$ and $\circ$, 
respectively. Data points are from Table\,\ref{txlum}, and for 47\,Tuc 
from Verbunt \&\ Hasinger (1998) slightly modified for an assumed 0.6\,keV
bremsstrahlung spectrum. In all cases the detection limits in and outside the
core are close to the lowest detected luminosities in and outside the cores.
\label{fxlum}}
\end{figure}

We can interpret the ROSAT and ASCA observations in two ways. The first
and most likely is that ASCA indeed did detect the Rapid Burster in quiescence,
or another low-luminosity source in Liller\,1; and that ROSAT observed
when this source had a lower flux level. In fact, variation of transients
in their quiescent state is common (e.g.\ Campana et al.\ 1997).
The star detected with ROSAT in this case is not detected with ASCA,
presumably because its spectrum is too soft. 
The second interpretation is that ASCA in fact detected the star also
detected with ROSAT, and not the quiescent counterpart of the Rapid Burster.
The position of the ROSAT source is marginally compatible with that
of the ASCA source; its countrate is exactly that predicted on the
basis of the ASCA source.
\nocite{cmsc97}
Dr.\ Asai has kindly communicated a new determination of the position of the
X-ray source detected by ASCA, using new calibrations to improve the
accuracy. This position, listed in Table~\ref{tabpos}, excludes
the ROSAT source as a possible counterpart, and thus confirms that ASCA
indeed detected a source in the cluster.

\section{Summary and discussion}

In the three low-reddened clusters \omcen, NGC\,6397 and NGC\,6752 we have
detected a total of 17 dim X-ray sources, of which 5 are well outside
the core. The X-ray luminosities of these sources are listed in
Table\,\ref{txlum}, and plotted in Fig.\,\ref{fxlum}.
The interpretation of Fig.\,\ref{fxlum} must be made with some care.
First, sources outside the core may not belong to the cluster; 
the faintest core source in \omcen\ may be a fore- or background source.
Second, the conversion of observed countrate
to luminosity depends on the assumed spectrum,
and from PSPC observations we know that different sources
have different spectral parameters (Johnston et al.\ 1994).
For example, the 0.6\,keV black
body spectrum used for the sources in \omcen\ gives a 40\%\ higher
flux for the same countrate than an assumed 0.6\,keV bremsstrahlung
spectrum would give. The bremsstrahlung spectrum is used for the three
other clusters.
Third, the detection limits in NGC\,6397,
NGC\,6752 and \object{47~Tuc} are higher in the cores, where the point spread
functions of sources overlap, than outside the core.
Such a difference is not present in \omcen.
Fourth, we show the average luminosity, and several sources are known to
be variable.

\begin{table}
\caption{X-ray luminosities in \ergs\ in the 0.5-2.5 keV band 
of the dim X-ray sources in globular clusters described in this paper. 
For sources in \omcen\ we assume a 0.6\,keV black body spectrum; 
for those in NGC\,6397 and NGC\,6752 a 0.6\,keV bremsstrahlung spectrum. 
For the same countrate, the blackbody spectrum corresponds to a flux higher 
by about 40\%\ than the bremsstrahlung spectrum.
\label{txlum}}
\begin{tabular}{rl@{\hspace{1.cm}}rl@{\hspace{1.cm}}rl}
\multicolumn{2}{l}{\omcen} & \multicolumn{2}{l}{NGC\,6397} &
\multicolumn{2}{l}{NGC\,6752} \\
X & $\log L_{\rm x}$ & X & $\log L_{\rm x}$ & X & $\log L_{\rm x}$ \\
\multicolumn{2}{c}{core} & \multicolumn{2}{c}{core} & \multicolumn{2}{c}{core}
 \\
9a & 32.14 &          13 & 31.84      & 7a & 31.86 \\
9b & 32.22 &          4a & 31.73      & 7b & 31.89 \\
20 & 31.96 &          4b & 31.75      & 21 & 31.35 \\
\multicolumn{2}{c}{outside} & 4c & 31.43 & 22 & 31.26 \\
 7 & 32.33 &           4d & 31.53     & \multicolumn{2}{c}{outside} \\
21 & 31.88 & \multicolumn{2}{c}{outside} & 6 & 31.66  \\
   &       &          12 & 30.95       & 14 & 31.50 \\
\end{tabular}
\end{table}

With these points in mind, we note from Fig.\,\ref{fxlum} that in all
clusters except possibly \omcen\ the most luminous sources appear to be in 
the cluster core.
The main difference between \omcen\ and the other clusters is that
the collision frequency in \omcen\ is so low that one expects no low-mass
X-ray binaries in it, and that most cataclysmic variables in it will
be evolved from primordial binaries (Verbunt \&\ Meylan 1988, Davies 1997).
In addition, the mass segregation in this cluster is very low.
Thus in \omcen\ there is no marked difference between the core and the
regions outside the core.
\nocite{dav97}\nocite{vm88}

In each cluster we detect sources down to the detection limit; this suggests
that more sensitive observations will detect more sources. 
In the cores of NGC\,6397 and NGC\,6752 the detection of more source will 
also require better imaging, so that the faint sources can be detected against 
the brighter ones.
We do not detect a difference between the luminosities of sources in the
collapsed globular cluster NGC\,6397 and the much less concentrated globular 
cluster NGC\,6752.
On the other hand, the highly concentrated cluster 47~Tuc contains three
sources which are an order of magnitude brighter than the brightest
sources in NGC\,6397 and NGC\,6752.

\begin{figure}
%\centerline{\psfig{figure=compa.ps,width=0.7\columnwidth,clip=t} {\hfil}}
\centerline{\psfig{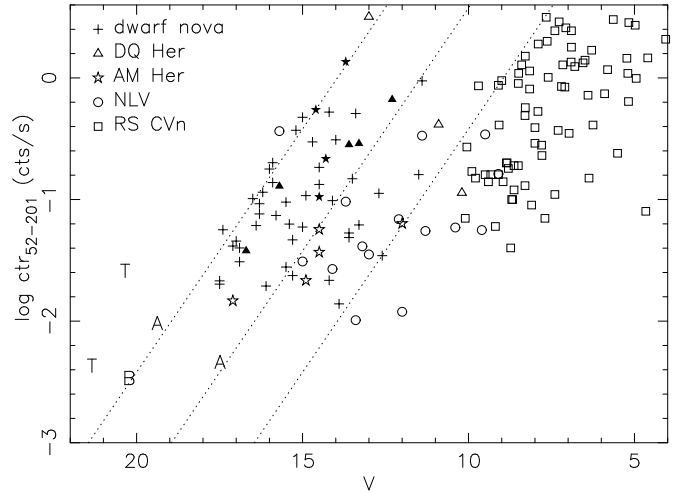} {\hfil}}
\caption{X-ray countrates of the dim sources in globular clusters as
a function of their visual magnitude, compared with the ROSAT PSPC
countrates and visual magnitudes of various types of
cataclysmic variables (data from Verbunt et al.\ 1997;
filled symbols represent systems first discovered in X-rays
and only subsequently identified as cataclysmic variables,
i.e.\ X-ray selected systems) and with RS CVn systems
(data from Dempsey et al.\ 1993) respectively. 
PSPC countrates of the dim cluster sources have been computed for an
assumed 0.6\,keV bremsstrahlung spectrum, corrected for absorption, from
the observed HRI countrates. Visual magnitudes are also corrected for 
absorption. T indicate sources in 47\,Tuc (\x9 and \x19, $V$ as estimated by
Verbunt \&\ Hasinger 1998), A in NGC\,6397 (\x4b and \x4c, $V$ from Cool et 
al.\ 1998), B in NGC\,6752 (\x7a, $V$ from Bailyn et al.\ 1996).
The dotted lines indicate a constant ratio of X-ray to optical flux.
\label{fcompa}}
\end{figure}
\nocite{dlfs93}

Viable optical counterparts have been suggested for only five among the
26 sources shown in Fig.\,\ref{fxlum}, all of them probable cataclysmic 
variables.
We compare the ratio of X-ray flux to optical flux of these sources
with the ratios measured for cataclysmic variables and for RS CVn systems
in the Galactic Disk in Fig.\,\ref{fcompa}.
It is seen that the suggested optical counterparts for the sources in 
NGC\,6397 and NGC\,6752 lead to ratios which
are compatible with those of cataclysmic variables, whereas those in 47~Tuc
are too bright in X-rays, in agreement with Fig.\,\ref{fxlum}. If these
sources are indeed cataclysmic variables, their excessive X-ray luminosity
needs to be explained; alternatively, the suggested identifications
may be chance coincidences (as discussed by Verbunt \&\ Hasinger 1998).
All suggested counterparts lead to higher X-ray to optical flux ratios
than those of RS CVn binaries.

The accurate positions that we determine for individual sources are valid
for separately detected sources in particular. In the case of overlapping
sources, we do not have unique solutions. Thus, in the core of NGC\,6397
fits with 5 and 6 sources are both acceptable, at similar quality; and we 
cannot exclude that more sources contribute to the observed flux, which
would invalidate our derived positions.

Binaries may reside away from the core either because the cluster
has undergone little mass segregation, or because a three-body interaction
(i.e.\ a close encounter of a binary with a single star) in the core has 
expelled the binary from the core (e.g.\ Hut et al.\ 1992). \nocite{hmr92}
In the latter case the binary is expected
to be eccentric immediately after being expelled; tidal forces may in time
circularize the orbit again. 
Such binaries are only a minority of the overall binary population of
a cluster; however, X-ray observations may preferably select such
binaries if tidal forces act in them.
Since sources away from the core can be
fore- or background sources, optical identification of them is required to
settle whether they belong to the cluster or not.
Our accurate positions should help in finding such counterparts.

\begin{acknowledgements}
We have made use of the ROSAT Data Archive of the Max Planck
Institut f\"ur extraterrestrische Physik at Garching; of the SIMBAD database
operated at Centre de Donn\'ees astronomiques in Strasbourg; and of the 
Digitized Sky Surveys. For Hipparcos and Tycho data the CD-ROM Celestia, 
provided by ESA, was very helpful. 
We further thank Lucien Kuiper for discussions about the Maximum Likelihood
algorithms, Marten van Kerkwijk for help with
use of the Digitized Sky Surveys, and Dr. Asai for communication of the
re-determination of the position of the ASCA source in Liller\,1.
\end{acknowledgements}

\end{document}